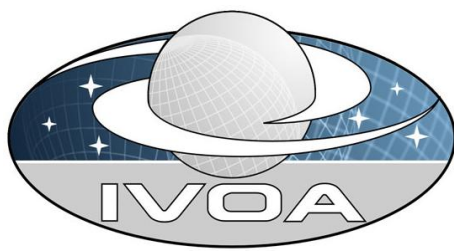

# IVOA Single-Sign-On Profile: Authentication Mechanisms

# Version 2.0

## IVOA Recommendation 2017-05-24

Working group
: http://www.ivoa.net/twiki/bin/view/IVOA/IvoaGridAndWebServices

This version
: http://www.ivoa.net/documents/SSOAuthMech/20170524

Latest version
: http://www.ivoa.net/documents/SSOAuthMech

Previous versions
: WD-20160930
  REC-1.01
  PR-20070904
  PR-20070621
  WD-20060519

Author(s)
: Giuliano Taffoni, André Schaaff, Guy Rixon, Brian Major

Editor(s)
: Giuliano Taffoni

## Abstract

Approved client-server authentication mechanisms are described for the IVOA single-sign-on profile: No Authentication; HTTP Basic Authentication; TLS with passwords; TLS with client certificates; Cookies; Open Authentication; Security Assertion Markup Language; OpenID. Normative rules are given for the implementation of these mechanisms, mainly by reference to pre-existing standards. The Authorization mechanisms are out of the scope of this document.

## Status of This Document

This document has been reviewed by IVOA Members and other interested parties, and has been endorsed by the IVOA Executive Committee as an IVOA Recommendation. It is a stable document and may be used as reference material or cited as a normative reference from another document. IVOA's role in making the Recommendation is to draw attention to the specification and to promote its widespread deployment. This enhances the functionality and interoperability inside the Astronomical Community.

A list of current IVOA Recommendations and other technical documents can be found at http://www.ivoa.net/documents/.

## Contents

# Acknowledgments

This document derives from discussions among the Grid and Web Services working-group of IVOA. It is particularly informed by prototypes built by Matthew Graham (Caltech/US-NVO), Paul Harrison (ESO/EuroVO), David Morris (Cambridge/AstroGrid), Raymond Plante (NCSA/US-NVO) Brian Major and Donovan Patrick Dowler (CADC) and Giuliano Taffoni (INAF-VObs.it). The prior art for the use of proxy certificates comes from the Globus Alliance. This document has been developed with support from the National Science Foundation's Information Technology Research Program with The Johns Hopkins University, from the UK Particle Physics and Astronomy Research Council (PPARC) and from the European Commission's Work programme FP7 via the CoSADIE project and the H2020 via the ASTERICS project.

# Conformance-related definitions

The words "MUST", "SHALL", "SHOULD", "MAY", "RECOMMENDED", and "OPTIONAL" (in upper or lower case) used in this document are to be interpreted as described in IETF standard, Bradner (1997).

The *Virtual Observatory (VO)* is general term for a collection of federated resources that can be used to conduct astronomical research, education, and outreach. The International Virtual Observatory Alliance (IVOA) is a global collaboration of separately funded projects to develop standards and infrastructure that enable VO applications.

# 1 Introduction

IVOA's single-sign-on architecture is a system in which users assign cryptographic credentials to user agents so that the agents may act with the user's identity and access rights. This standard describes how agents use those credentials to authenticate the user's identity in requests to services. This standard describes also the authentication mechanism of an application or a service making a call (on behalf of someone or something else) to an API or to another service. This document is essentially a *profile* against existing security standards; that is, it describes how an existing standard should be applied in an IVOA application to support single sign-on capabilities in the IVOA. In the following sections, we make specific references to details spelled out in these standards. For the purposes of validating against this standard, those referenced documents should be consulted for a full explanation of those details. Unfortunately, a reader that is unfamiliar with these external standards might find this specification confusing. To alleviate this problem, each major section is concluded by a Commentary subsection that provides some explanations of the detailed terms and concepts being referred to. The Commentary subsection may also provide recommended scenarios for how this specification might actually be realised. Note that the statements in the Commentary subsections are non-normative and should not be considered part of precise specification; nevertheless, they are indicative of the intended spirit of this document.

## 1.1 Role within the VO Architecture

Fig. 1 shows the role this document plays within the IVOA architecture (Arviset and Gaudet et al., 2010).

# 2 Scope of this standard

## 2.1 Requirements

When a service is registered in an IVOA registry, that service's resource document MAY include metadata expressing conformance to one or more of the authentication mechanisms approved in the IVOA SSO profile. Such a service MUST implement those mechanisms as described in this document, and clients of the service MUST participate in the mechanism when calling the service. If a service does not provide any SSO specification it is assumed that no authentication is required. The registration of the service interface SHALL contain an XML element of type `SecurityMethod` as specified in the XML schema for VOResource (Plante and Benson et al., 2008). The value of this element distinguishes the authentication mechanism using the values stated in the sections below. Services registered without the metadata alluded to above need not support any authentication mechanism. If they do require authentication, they MAY use either the IVOA-standard mechanisms or others that are not IVOA standards, but they MUST specify a `SecurityMethod` element.

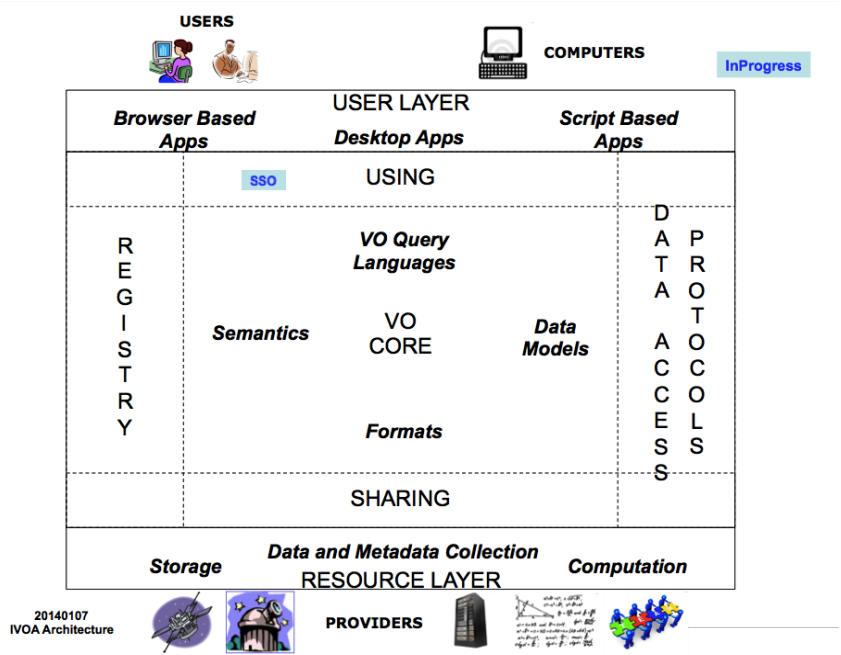

*Figure 1:* Architecture diagram for this document

## 2.2 Commentary

The IVOA SSO profile allows the development of a "realm" of interoperable services and clients. Service providers opt in to this realm by implementing this current standard and by registering accordingly in the IVOA registry. This allows clients to discover a secured service through the registry and to be able to use it without being customized for the details of the specific service.

Services within the Virtual Observatory that are not intended to be widely interoperable need not opt in to the SSO realm. In particular, "pri-vate" services, accessed by web browsers and protected by passwords, are allowed. However, these private services SHOULD be reworked to follow the IVOA standard if they are later promoted to a wider audience.

An example of a registration for a secured interface follows.

```
<interface xmlns:vs='ivo://www.ivoa.net/xml/VODataService/v1.1'
        xsi:type='vs:ParamHTTP'>
    <accessURL>http://some.where/some/thing</accessURL>
    <securityMethod>ivo://ivoa.net/sso#saml2.0</securityMethod>
</interface>
```

More than one `securityMethod` can be specified:

```
<interface xmlns:vs='ivo://www.ivoa.net/xml/VODataService/v1.1'
        xsi:type='vs:ParamHTTP'>
    <accessURL>http://some.where/some/thing</accessURL>
    <securityMethod>ivo://ivoa.net/sso#saml2.0</securityMethod>
    <securityMethod>ivo://ivoa.net/sso#cookie</securityMethod>
    <securityMethod>ivo://ivoa.net/sso#OpenID</securityMethod>
</interface>
```

The order of the `securityMethod` elements determines the priority of the method to use. In the example above, the preferred method to access the service is *SAML*, then *cookies*, and finally, if the others are not available, *OpenID*.

## 3  Approved authentication mechanisms

The following authentication mechanisms are approved for use in the SSO profile.

- No authentication required.
- HTTP Basic Authentication
- Transport Layer Security (TLS) with passwords.
- Transport Layer Security (TLS) with client certificates.
- Cookies
- Open Authentication (OAuth)
- Security Assertion Markup Language (SAML)
- OpenID

The mechanism is associated with the interface provided by the service and registered in the IVOA registry.

Services that are registered with a IVOA registry as having a *WebService* type interface (as described in the VOResource document) SHALL support OAuth, or SHALL support cookies or SHALL support TLS with client certificates or SHALL require no authentication. Interfaces by which a user logs in to the SSO system SHALL support either TLS with client certificates, or TLS with passwords, or SAML or a combination of them.

## 3.1 List of approved authentication mechanisms and the corresponding securityMethod

The approved authentication mechanisms and the corresponding `securityMethod` to implement is listed in the table below.

| SSO mechanism | `<securityMethod>` |
| --- | --- |
| HTTP Basic Authentication | `ivo://ivoa.net/sso#BasicAA` |
| TLS with password | `ivo://ivoa.net/sso#tls-with-password` |
| TLS with client certificate | `ivo://ivoa.net/sso#tls-with-certificate` |
| Cookies | `ivo://ivoa.net/sso#cookie` |
| Open Authentication | `ivo://ivoa.net/sso#OAuth` |
| SAML | `ivo://ivoa.net/sso#saml2.0` |
| OpenID | `ivo://ivoa.net/sso#OpenID` |

# 4 HTTP Basic Authentication

## 4.1 Requirements

Services using HTTP basic authentication SHALL use the authentication mechanism described in the RFC7235 (Fielding, 2014) that updates RFC2617 (Franks and Hallam-Baker et al., 1999). Interfaces using this mechanism SHALL be registered with the security method
    `ivo://ivoa.net/sso#BasicAA`

## 4.2 Commentary

HTTP provides a simple challenge-response authentication framework that can be used by a server to challenge a client request and by a client to provide authentication information. The HTTP authentication framework does not define a single mechanism for maintaining the confidentiality of credentials. HTTP depends on the security properties of the underlying transport or session-level connection to provide confidential transmission of header fields. Connection secured with TLS are RECOMMENDED prior to exchanging any credentials.

The "HTTP basic authentication" SHOULD be used with particular attention as sensible information (password) are sent over the wire in base64 encoding (which can be easily converted to plaintext) exposing the user to the possibility her credentials to be stolen.

## 5 Details of TLS

### 5.1 Requirements

Services using Transport Layer Security (TLS) SHALL do so according to the TLS v1.2 standard RFC5246 (Dierks and Rescorla, 2008).

### 5.2 Commentary

TLS supersedes the Secure Sockets Layer that is an outdated cryptographic protocol. TLS v1.0 was based on SSL v3.0; the current version of TLS is V1.2 described in by Dierks and Rescorla (2008). TLS v1.2 is backwards compatible with TLS v1.0, TLS v1.1 and SSL v3.0. "TLS versions 1.0, 1.1, and 1.2, and SSL 3.0 are very similar, and use compatible ClientHello messages; thus, supporting all of them is relatively easy.[...] TLS 1.2 clients that wish to support SSL 2.0 servers MUST send version 2.0 CLIENT-HELLO messages defined in SSL2." (Dierks and Rescorla, 2008).

## 6 Details of TLS-with-client-certificate

### 6.1 Requirements

Certificates SHALL be transmitted and checked according to the TLS v1.2 standard RFC5246.

Services implementing TLS MUST support certificate chains including proxy certificates according to RFC6818 (Yee, 2013).

Interfaces using this mechanism SHALL be registered with the security method
    `ivo://ivoa.net/sso#tls-with-certificate`

### 6.2 Commentary

When Mutual Certificate Authentication is configured for REST services, both the client and the service perform identity verification or authentication through X.509 certificates.

The client authenticates the service during the initial SSL handshake, when the server sends the client a certificate to authenticate itself.

# 7 Details of TLS-with-password

## 7.1 Requirements

The user-name and password SHALL be passed in the message protected by the TLS mechanism, not as part of the mechanism itself.

Interfaces using this mechanism SHALL be registered with the security method
```
ivo://ivoa.net/sso#tls-with-password
```

## 7.2 Commentary

"HTTP basic authentication" passes the user-name and password in the HTTP headers, assuming that the credentials are not a natural part of the message body. This standard applies the TLS-with-Password mechanism only to the special case of logging in to the SSO realm. Hence, the user-name and password are logically part of the message body, not the message header.

# 8 The use of Cookies

## 8.1 Requirements

Cookie-Based Authentication uses server side cookies to authenticate the user on every request. The way to manage cookies for authentication is described in RFC6265 (Barth, 2013).

Interfaces using this mechanism SHALL be registered with the security method
```
ivo://ivoa.net/sso#cookie
```

## 8.2 Commentary

RESTful web services MAY support session-based authentication, either by establishing a session token via a POST or by using an API key as a POST body argument or as a cookie. User-names, passwords, session tokens, and API keys SHOULD not appear in the URL, as this can be captured in web server logs, which makes them intrinsically valuable.

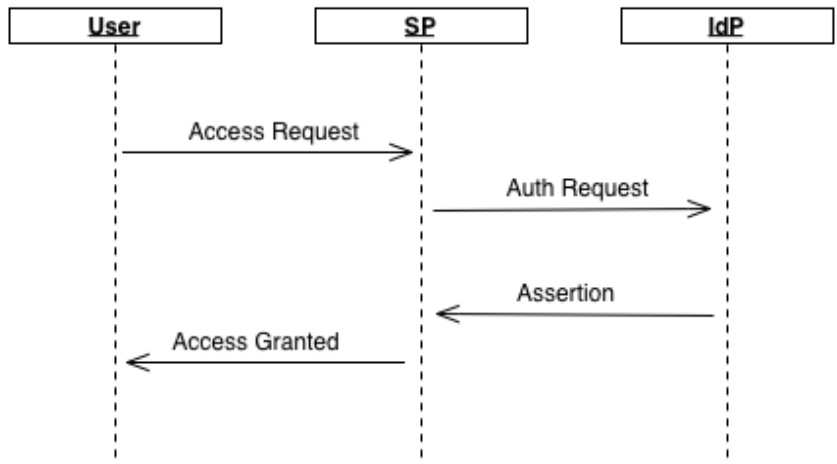

*Figure 2:* Simplified picture of SAML 2.0 authentication.

# 9 Details on SAML authentication

## 9.1 Requirements

Services using SAML authentication mechanisms SHALL do so according to the saml-core-2.0-os OASIS standard (Cantor and Kemp et al., 2005*a*). SAML includes protocols and protocol bindings and security (Cantor and Kemp et al., 2005*b*).

Interfaces using this mechanism SHALL be registered with the security method

    ivo://ivoa.net/sso#saml2.0

## 9.2 Commentary

SAML presumes two primary roles in any transaction: the organisation where the identity is established, known as the Identity Provider ("IdP"), or Asserting Party ("AP"); and the organisation which (for this transaction) wants to use this identity, known as the Service Provider ("SP"), or Relying Party ("RP").

A user attempts to access an application with the Service Provider. The SP needs to establish the identity of this user, and so sends an authentication request to the Identity Provider.

The user authenticates with the IdP (IdP is taking care of the authentication mechanisms and protocols e.g. Kerberos, ldap etc.) so the IdP can send back an 'Assertion' to the SP. Now the SP knows who the user is, and can process that user accordingly (see Fig. 2).

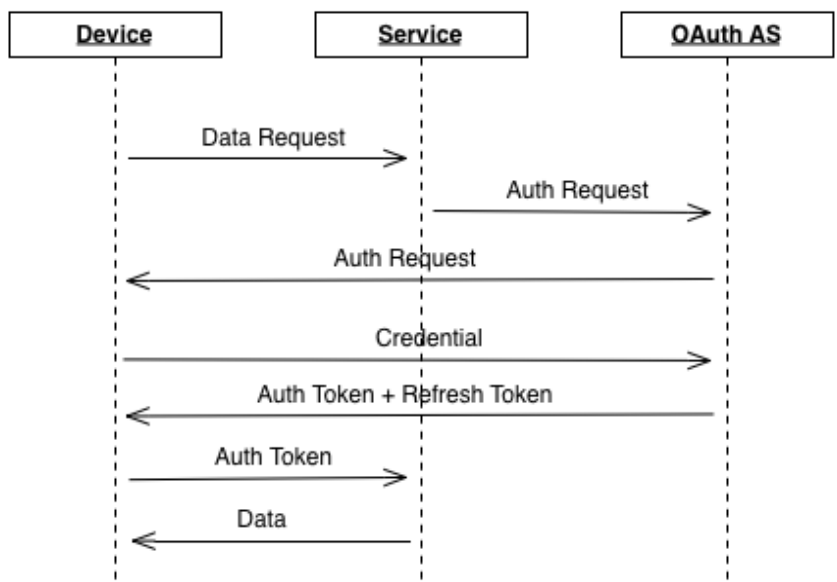

*Figure 3:* Simplified picture of OAuth 2.0 authentication.

SAML2.0 protocol allows also to implement authentication service discovery mechanisms. SAML2.0 defines a browser-based protocol by which a centralized discovery service can provide a requesting service provider with the unique identifier of an IdP that can authenticate the user.

## 10 Details on OAuth

### 10.1 Requirements

Services using OAuth authentication mechanisms SHALL do so according to the RFC6749 (Hardt, 2012).

Interfaces using this mechanism SHALL be registered with the security method
```
ivo://ivoa.net/sso#OAuth
```

### 10.2 Commentary

Open Authentication 2.0 (also in conjunction with OpenID Connect) is actually the adopted standard to handle identity in the framework of RESTful web services. OAuth is used when an application is making a request on behalf of a user.

OAuth introduces the notion of an 'authorization token', a 'refresh token' and Authorization Service (AS). The 'authorization' token states that the client application has the right to access services on the server (see Fig. 3). However, it does not supersede any access control decisions that the server-side application might make.

OAuth protocol can be implemented to delegate credential from an application to another.

## 11 Details on OpenID

### 11.1 Requirements

Services using OpenID authentication mechanisms SHALL do so according to the OpenID Foundation standards (OpenID, 2007)

Interfaces using this mechanism SHALL be registered with the security method

```
ivo://ivoa.net/sso#OpenID
```

### 11.2 Commentary

OpenID is an open and decentralized authentication and identity system. OpenID relying parties do not manage end user credentials such as passwords or any other sensitive information which makes authentication and identity management much simpler and secure. In a RESTful environment OpenID Connect (Sakimura and Bradley et al., 2014) is commonly adopted as authentication solution. "OpenID Connect is a simple identity layer on top of the OAuth 2.0 protocol, which allows computing clients to verify the identity of an end-user based on the authentication performed by an authorization server, as well as to obtain basic profile information about the end-user in an interoperable and REST-like manner." (OpenID, 2007).

## 12 Conclusions

This document presents a list of security standards that may be implemented when developing a service that requires authentication. The list includes the most frequently used standards at the time this document has been produced.

In this document we are presenting two types of SSO protocols: "local" and "federated". Local SSO provides solutions for keeping a repository of user-names and passwords that could be used transparently across several internal applications but it is local to one domain/service.

Federated identity means linking and using the electronic identities a user has across several identity management systems. In simpler terms, a service does not necessarily need to obtain and store users credentials in order to authenticate them. Instead, the service (or the application) can use an identity management system that is already storing a user's electronic identity to authenticate the users given, of course, that the application trusts that identity management system. Federated identities are convenient for users, since they don't have to keep a set of user-names and passwords for every single application that they use and for service providers that do not need to store and manage credentials.

Local SSO is managed by the following protocols: HTTP Basic Authentication, Transport Layer Security (TLS) with passwords, cookies OAuth, SAML, OpenID and Transport Layer Security (TLS) with client certificates (thanks to the CA trust) are protocol that allow to implement federated SSO.

The choice the authentication to use is related to the project/service requirements, we suggest at least to implement a local authentication based on Transport Layer Security (TLS) with passwords, that allows a reasonable security framework for exchanging authentication tokens.

More complex projects/services that need to offer resources to large communities should prefer federated identities. For example SAML2.0 is the protocol used to build the EduGain World wide identity federation for education and research.

## A  VOResource SecurityMethod

This Appendix presents an extract of the VOResource Description XML schema. Here we present the part of the schema regarding the `SecurityMethod` element to facilitate the reader identify the relevant schema sections in the VOResource Description.

```
<xs:schema xmlns="http://www.w3.org/2001/XMLSchema"
xmlns:xs="http://www.w3.org/2001/XMLSchema"
xmlns:vr="http://www.ivoa.net/xml/VOResource/v1.0"
xmlns:vm="http://www.ivoa.net/xml/VOMetadata/v0.1"
targetNamespace="http://www.ivoa.net/xml/VOResource/v1.0"
elementFormDefault="unqualified" attributeFormDefault="unqualified" version="1.02">
<xs:annotation>...</xs:annotation>
<xs:simpleType name="UTCTimestamp">...</xs:simpleType>
<xs:simpleType name="UTCDateTime">...</xs:simpleType>
<xs:complexType name="Resource">...</xs:complexType>
<xs:simpleType name="ValidationLevel">...</xs:simpleType>
<xs:complexType name="Validation">...</xs:complexType>
<xs:simpleType name="AuthorityID">...</xs:simpleType>
<xs:simpleType name="ResourceKey">...</xs:simpleType>
<xs:simpleType name="IdentifierURI">...</xs:simpleType>
<xs:simpleType name="ShortName">...</xs:simpleType>
<xs:complexType name="Curation">...</xs:complexType>
<xs:complexType name="ResourceName">...</xs:complexType>
<xs:complexType name="Contact">...</xs:complexType>
<xs:complexType name="Creator">...</xs:complexType>
<xs:complexType name="Date">...</xs:complexType>
<xs:complexType name="Content">...</xs:complexType>
```

```xml
<xs:complexType name="Source">...</xs:complexType>
<xs:simpleType name="Type">...</xs:simpleType>
<xs:simpleType name="ContentLevel">...</xs:simpleType>
<xs:complexType name="Relationship">...</xs:complexType>
<xs:complexType name="Organisation">...</xs:complexType>
<xs:complexType name="Service">...</xs:complexType>
<xs:simpleType name="Rights">...</xs:simpleType>
<xs:complexType name="Capability">...</xs:complexType>
<xs:complexType name="Interface" abstract="true">
<xs:annotation>...</xs:annotation>
<xs:sequence>
   <xs:element name="accessURL" type="vr:AccessURL"
            minOccurs="1" maxOccurs="unbounded">...</xs:element>
   <xs:element name="securityMethod" type="vr:SecurityMethod"
            minOccurs="0" maxOccurs="unbounded">
     <xs:annotation>
        <xs:documentation> the mechanism the client must employ to
                                 gain secure access to the service.
        </xs:documentation>
        <xs:documentation> when more than one method is listed, each one
                                 must be employed to gain access.
        </xs:documentation>
     </xs:annotation>
   </xs:element>
</xs:sequence>
<xs:attribute name="version" type="xs:string" default="1.0">...</xs:attribute>
<xs:attribute name="role" type="xs:NMTOKEN">...</xs:attribute>
</xs:complexType>
<xs:complexType name="AccessURL">...</xs:complexType>
<xs:complexType name="SecurityMethod">
   <xs:annotation>
      <xs:documentation>a description of a security mechanism.</xs:documentation>
      <xs:documentation> this type only allows one to refer to the mechanism via a URI.
                              Derived types would allow for more metadata.
      </xs:documentation>
   </xs:annotation>
<xs:sequence/>
<xs:attribute name="standardID" type="xs:anyURI">
   <xs:annotation>
      <xs:documentation> A URI identifier for a standard security mechanism. </xs:documentation>
      <xs:documentation>
          This provides a unique way to refer to a security specification standard.
          The use of an IVOA identifier here implies that a VOResource
          description of the standard is registered and accessible.
      </xs:documentation>
   </xs:annotation>
</xs:attribute>
</xs:complexType>
<xs:complexType name="WebBrowser">...</xs:complexType>
<xs:complexType name="WebService">...</xs:complexType>
</xs:schema>
```

# B Changes from Previous Versions

## B.1 Changes from v. 1.01

- We remove all the references to SOAP as deprecated from IVOA
- We add new security methods and relative discussion sessions: OpenID, SAML, Cookies, HTTP basic authentication